# Brushless Motor Performance Optimization by Eagle Strategy with Firefly and PSO

Appalabathula Venkatesh[1], Pradeepa H[*2], Chidanandappa R[3], Shankar Nalinakshan[4], Jayasankar V N[5]

[1]*Research Scholar, Electrical and Electronics Engineering, The National Institute of Engineering, Mysuru, Karnataka, India.*

[*2]*Associate Professor, Electrical and Electronics Engineering, The National Institute of Engineering, Mysuru, Karnataka, India.*

[3]*Associate Professor, Electrical and Electronics Engineering, The National Institute of Engineering, Mysuru, Karnataka, India.*

[4]*Assistant Professor, Electrical and Electronics Engineering, The National Institute of Engineering, Mysuru, Karnataka, India.*

[5]*Assistant Professor, Electrical and Electronics Engineering, The National Institute of Engineering, Mysuru, Karnataka, India.*

[1]venkatesh@nie.ac.in, [2] pradeep3080@nie.ac.in, [3] chidananda@nie.ac.in, [4] shankar.nalinakshan@nie.ac.in, 5 jayasankarvn@nie.ac.in

***Abstract** — Brushless motors has special place though different motors are available because of its special features like absence in commutation, reduced noise and longer lifetime etc., The experimental parameter tracking of BLDC Motor can be achieved by developing a Reference system and their stability is guaranteed by adopting Lyapunov Stability theorems. But the stability is guaranteed only if the adaptive system is incorporated with the powerful and efficient optimization techniques. In this paper the powerful eagle strategy with Particle Swarm optimization and Firefly algorithms are applied to evaluate the performance of brushless motor Where, Eagle Strategy(ES) with the use of Levy's walk distribution function performs diversified global search and the Particle Swarm Optimization (PSO) and Firefly Algorithm(FFA) performs the efficient intensive local search. The combined operation makes the overall optimization technique as much convenient The simulation results are obtained by using MATLAB Simulink software.*

## I. INTRODUCTION

To investigate the Linear Time Invarient (LTI) systems conventional techniques like the linear quadratic regulator (LQR), pole-placement technique and MRAS are helpful in stabilizing the system where control design problems doesn't concern about the contrary objectives.

In general, While designing or adapting the PID controllers, a compromise has to be undertaken between performance and robust stability. In LQR design,Due to feedback presence-one has to consider the diminishing of the process disturbances and the oscillations created by measurement noise which is subjected into the plant or system. But the Direct MRAS has the ability to provide a desirable solution to mitigate the tracking errors [1].

When there are too many variable or effective plant parameters are presented in the system. In such case, Direct MRAS control algorithm may produce undesirable solutions to such robust systems. For such a robust systems- indirect MRAS has the ability in providing an effective solution, to enhance the control strategies but sometimes because of the presence of parametric uncertainties, every time it may not be possible to get the reduction in tracking errors.

For such cases, where more parametric variations are presented there Linear control techniques such as the full-state feedback was tested but had no success in controlling both a balancing angle and a position of the Mobile Robots explained in [2], [3]

However, Intelligent control techniques like NNs and Fuzzy Logic Controllers(FLCS) take much time in framing the mapping/Rule based conditions to obtain the desired rules and conditions, even though these techniques having the ability to give better results but time concerns more as a negative features [3]-[4].

In the model reference adaptive system the desired closed loop response is specified through a stable reference model. The control system attempts to make the process output similar to the reference model output [5]-[6]. By the application of the standard Lyapunov stability theorems along with the Model and Plant one can overcome the drawbacks which are mentioned previously with MRAS.

In practical there is always an uncertainty, noise signals which plays key role in performance outcome of a systems. These disturbances we cannot eliminate completely but we can reduce their effects. The stability is guaranteed only if the adaptive system is incorporated with the powerful and efficient optimization techniques. The metaheuristic powerful optimization techniques like eagle strategy with PSO[7]-[8] and FFA which are from natures inspiration, these methods are applied to evaluate the performance of brushless motor. The following Figure 1 shows the evaluation of BLDC Motor over other AC and DC motors. The special motors 4,5 which are derived from the motors 1,2,3 by replacing field windings with permanent magnets. The synchronous motor(SM) immediately becomes brushless, but the d.c. motor must go through an additional transformation, from motors 4,5 to motors 6,7 with the inversion of the stator and rotor, before the brushless version is achieved[9].

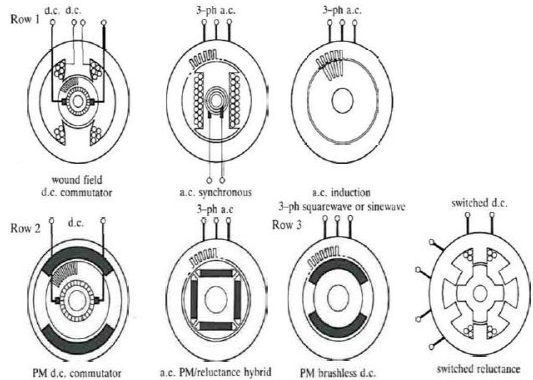

**Fig. 1** Evaluation of BLDC motor from conventional AC and DC Motors

## II. MATHEMATICAL MODELLING

### A. Modelling of MRAS

An adaptive system is an extension of the basic feedback structured LQR system, The developed adaptive controllers are integrated with the basic control systems with parametric uncertainties are taken into account along with the internal, external disturbances to the plant, which controls the plant in achieving desired performances with the help of desired stable model.

Let, the plant transfer function with process parameters is as follows

$$\frac{Y_p(s)}{U(s)} = \frac{b_p}{s^2 + a_p s}$$

$$s^2 Y_p(s) + a_p s Y_p(s) = b_p(s) U(s)$$

Let

$$Y_p(t) = x_{1p}(t)$$

$$\Rightarrow \dot{Y}_p(t) = \dot{x}_{1p} = x_{2p}$$

$$\Rightarrow \ddot{Y}_p(t) = \ddot{x}_{1p} = \dot{x}_{2p}$$

$$\dot{x}_{2p} = b_p u(t) - a_p x_{2p}(t)$$

Hence state space equations for a plant are

$$\dot{x}_{1p} = x_{2p}(t)$$

$$\dot{x}_{2p} = b_p u(t) - a_p x_{2p}(t)$$

Let the reference model transfer function

$$\frac{Y_m(s)}{U(s)} = \frac{\omega_n^2}{s^2 + 2\zeta\omega_n s}$$

$$s^2 Y_m(s) + 2\zeta\omega_n s Y_m(s) = \omega_n^2 U(s)$$

Let

$$Y_m(t) = x_{1m}(t)$$

$$\Rightarrow \dot{Y}_m(t) = \dot{x}_{1m} = x_{2m}$$

$$\Rightarrow \ddot{Y}_m(t) = \ddot{x}_{1m} = \dot{x}_{2m}$$

$$\dot{x}_{2m} = \omega_n^2 U(t) - 2\zeta\omega_n x_{2m}(t)$$

Hence state space equations for a reference model

$$\dot{x}_{1m} = x_{2m}(t)$$

$$\dot{x}_{2m} = \omega_n^2 U(t) - 2\zeta\omega_n x_{2m}(t)$$

Where variations in process parameters $b_p$ and $a_p$ are compensated by variations in $K_p$ and $K_d$

Add the two state variables

$$\varepsilon = (R - x_{1p}); x_{2p}$$

$$u(t) = K_p \varepsilon - K_d x_{2p}$$

Then modified Plant equations will be

$$\dot{x}_{1p} = x_{2p}(t)$$

$$\dot{x}_{2p} = b_p \left( K_p \varepsilon - K_d x_{2p} \right) - a_p x_{2p}(t)$$

$$\dot{x}_{2p} = b_p K_p \varepsilon - \left( a_p + K_d b_p \right) x_{2p}(t)$$

Similarly the modified reference model will be as follows

$$\dot{x}_{1m} = x_{2m}(t)$$

$$\dot{x}_{2m} = \omega_n^2 U(t) - 2\zeta\omega_n x_{2m}(t)$$

$$\dot{\varepsilon} = 0 - \dot{x}_{1m} = -x_{2m}$$

$$\dot{x}_{2m} = \omega_n^2 U(t) - 2\zeta\omega_n x_{2m}(t)$$

$$\begin{bmatrix} \dot{\varepsilon} \\ \dot{x}_{2m} \end{bmatrix} = \begin{bmatrix} 0 & -1 \\ \omega_n^2 & -2\zeta\omega_n \end{bmatrix} \begin{bmatrix} \varepsilon \\ x_{2m} \end{bmatrix}$$

$$x_m = \begin{bmatrix} \varepsilon_m \\ x_{2m} \end{bmatrix}$$

$$A_m = \begin{bmatrix} 0 & -1 \\ \omega_n^2 & -2\zeta\omega_n \end{bmatrix} \quad B_m = \begin{bmatrix} 0 \\ 0 \end{bmatrix}$$

State equation of Model

$$\dot{x}_m = A_m x_m + B_m U$$

Output equation of the model

$$y_m = C_m x_m + D_m U$$

**A. Modelling of BLDCM**

The Transfer Function modelling of BLDC Motor is expressed as follows.

$$\frac{\omega_m(s)}{V_d(s)} = \frac{K_t}{L_a s^2 + (R_a J + L_a B)s + (R_a B + K_e K_t)}$$

Where

$\tau_m = \dfrac{R_a J}{K_e K_t} \rightarrow$ Mechanical time constant

$\tau_e = \dfrac{L_a}{R_a} \rightarrow$ Electrical time constant

$\omega_m(s) \rightarrow$ Angular speed of BLDCM

$V_d(s) \rightarrow$ Reference Input applied to BLDCM

By substituting the above terms in Eqn(3) Then the modified Transfer Function of BLDC-Motor is expressed as follows in Eqn(4)

$$\frac{\omega_m(s)}{V_d(s)} = \frac{\left(\dfrac{1}{K_e}\right)}{\tau_m \tau_e s^2 + \tau_m s + 1} \quad (1)$$

## III. BLDC MOTOR WITH LYAPUNOV STABILITY CONDITIONS

For the stability analysis of any system either it may be a Linear or Non-Linear Lyapunov theorems are helpful which are explained in [10]

**For Linear Systems**

State space equation of a linear system is

$$\dot{x} = Ax$$

For the stability choose a objective function

$$V(x) = x^T Q x$$

Then differentiation of above function

$$\dot{V} = x^T Q \dot{x} + \dot{x}^T Q x = x^T P x \quad (2)$$

Where

$P = A^T Q + QA$ is known as the Lyapunov equation
For a Stable system P must be a positive definite suchthat $\dot{V} < 0$

**For Non-Linear Systems:**
State space equation of a Non- linear system is

$$\dot{x} = Ax + Bu$$
$$u = n(z)$$
$$z = C^T x$$

For the stability analysis of a Non-Linear System, choose a objective function which is of the form

$$V(x,z) = x^T Q x + \beta \int n(z) dz \quad (3)$$

For stability All eigen values of A should be lies on left half of s-plane the Non-Linearity confined to the region [0,K]
BLDCM with the application of Lyapunov Stability conditions is as follows, The Objective Function is

$$f(e_1, e_2) = e^T P e + a^T \alpha a + b^T \beta b \quad (4)$$

Where Alpha,Beta are the acceleration factors

$e_x, e_\phi \rightarrow$ Errors corresponding to BLDCM states

By Applying Lyapunov Theorems for attaining the stability of BLDC- Motor state estimation

i.e, $\dot{f}(e_1, e_2) < 0$

By Solving the above Stability conditions we can get the optimal conditions for stability.
Then the Controller Parameters are expressed as follows after applying the conditions of Lyapunov Stability Theorem[11].

$$K_p = \frac{K_e T_m T_e}{\alpha_{21}} \int (P_{21} e_1 + P_{22} e_2) \varepsilon \, dt + K_p(0)$$

$$K_d = \frac{-K_e T_m T_e}{\alpha_{22}} \int (P_{21} e_1 + P_{22} e_2) x_{2p} \, dt + K_{dx}(0)$$

The P Matrix is obtained as

$$A_m^T P + P A_m = -Q$$

Where, $P = \begin{bmatrix} P_{11} & P_{12} \\ P_{21} & P_{22} \end{bmatrix}$

## IV. BLDC MOTOR PERFORMANCE OPTIMIZATION WITH ES WITH PSO OR FIREFLY ALGORITHMS

Performance optimization of non-linear systems depends upon the powerful optimizing technique operation on objective function. ES with PSO (or) Fire Fly (FF) algorithm uses Levy's flight equation which is one of the powerful, efficient optimization techniques[12].

The control strategies involved in the ES-PSO algorithm is a two stage evaluation process which stabilises the non-linear systems behaviour along with the Lyapunov stability theorems in both local and global optimization constraints.

It gives the strong diversification in global constrained space(path) which makes the intensive local search which ultimately results into efficient optimization technique.

The following control strategy steps shows how the ES with PSO or FF optimization technique works with the Brushless DC motor performance evaluation.

**Step I**
Define the unknown control parameters and their values of BLDC motor.

**Step II**
Load the formulated objective function with the help of Lyapunov stability theorems which is of the form.

$$f(e_1, e_2) = e^T P e + a^T \alpha a + b^T \beta b$$

**Step III**
Randomly generate the initial swarm/firefly population and set the maximum no. of iteration count.

**Step IV**
Find state space model of the BLDC motor from its Transfer function model. Error vector in linear displacement form

$$e_x(k) = x_{1p} - x_{1m}$$

$$e_\theta(k) = x_{2p} - x_{2m}$$

Assign above obtained error vector which is defined in the objective function.

**Step V**
While
$\|e_x(k+1) - e_x(k)\|$ and $\|e_\theta(k+1) - e_\theta(k)\|$ are within the tolerance limits
Perform random global search with the help of Levys walk distribution function

$$Le(s, \lambda) \cong \frac{\lambda^2}{2} \left( \frac{\sin p}{p} \right) \frac{\Gamma(\lambda)}{s^{\lambda+1}} \quad (5)$$

Where $p = \dfrac{\pi \lambda}{2}$

$s \rightarrow$ Levys walking step length

$\Gamma(\lambda) \rightarrow$ Standard gamma function

$\lambda \rightarrow$ ranges from 1 to 3

If $\lambda = 1 \rightarrow$ characteristics of Stochastic tunnelling

$\lambda = 2 \rightarrow$ characteristics of Cauchy's distribution

$\lambda = 3 \rightarrow$ characteristics of Brownian's distribution

**Step VI**
Evaluate the objective function with the global best solution values which are obtained in the above step. The obtained values are helpful for the reference for the local search for the next stage which will be done with the help of either PSO or Fire Fly local search algorithms.
The following sections describes how local search is performed with PSO and FF algorithms.

*A. Local Search with PSO*
From the obtained global best values from the previous ES update the position and velocity equations.

$$x_i(k+1) = x_i(k) + v_i(k+1) \quad (6)$$

$$v_i(k+1) = w_i v_i(k) + c_1 r_1 \left( P_{bst_i} - x_i(k) \right) + c_2 r_2 \left( G_{bst_i} - x_i(k) \right)$$

Where,

$P_{bst_i} \rightarrow$ Personal/present best value

$G_{bst_i} \rightarrow$ Global best value from ES

$c_1, c_2 \rightarrow$ Correction factors which accelerates the optimization speed by controlling the swarm motion.

$r_1, r_2 \rightarrow$ random values

$w_i \rightarrow$ Weight of the ith swarm

Update the objective function with updated position and velocity

If $f(P_{bst_i}) < f(x_i(k))$

Then present value is the best optimum position for the corresponding ith swarm. If the condition is not satisfied iteration count is increased by 1 and the updated position and velocity are determined.
Plot the desired post simulation results.

*B. Local Search with FF Algorithm*
From the obtained global best values from the previous ES update the movement of i$^{th}$ Firefly towards the j$^{th}$ Firefly using the formula

$$x_i(k+1) = x_i(k) + \beta_i(k+1) \quad (7)$$

$$\beta_i(k+1) = \beta_0 \exp\left(-\gamma r_{ij}^2\right)\left(x_j(k) - x_i(k)\right) + \alpha\left(r - \frac{1}{2}\right) \quad (8)$$

Where,

$\beta_0 \rightarrow$ Initial attractiveness constant

$\gamma \rightarrow$ Absorption Coefficient

$r \rightarrow$ random value

$\alpha \rightarrow$ Correction factor which accelerates the optimization speed by controlling the Firefly motion.

$r_{ij} \rightarrow$ Distance between the fireflies

$$r_{ij} = \|x_i - x_j\|$$

Update the objective function with updated position and values

If $f(x_i(k+1)) < f(x_i(k))$

Then present value is the best optimum position for the corresponding ith firefly. If the condition is not satisfied iteration count is increased by 1 and the updated position is determined.
Plot the desired post simulation results.

## V. SIMULATION RESULTS

The following Table 1 represents the tuned values of PID controller gain values for the final iteration.

**TABLE I**
**TUNED PID CONTROLLER GAINS FOR ES WITH PSO AND FFA**

| Sl.No | PID Controller Gains | | |
|---|---|---|---|
| | $K_P$ | $K_I$ | $K_D$ |
| PSO | -112.180842 | 6.680 | 6.99 |
| FFA | -37.5588 | 58.1842 | -0.0134 |

The following simulation results represents the tuned values of ES with PSO.

swarm =

  1.0e+003 *

    0.0013    0.0013    0.0013    0.0013    0.0563    0.0563    1.0000
    0.0023    0.0023    0.0023    0.0023    0.0546    0.0546    1.0000
    0.0033    0.0033    0.0033    0.0033    0.0530    0.0530    1.0000
    0.0043    0.0043    0.0043    0.0043    0.0513    0.0513    1.0000
    0.0053    0.0053    0.0053    0.0053    0.0496    0.0496    1.0000
    0.0063    0.0063    0.0063    0.0063    0.0480    0.0480    1.0000
    0.0073    0.0073    0.0073    0.0073    0.0463    0.0463    1.0000
    0.0083    0.0083    0.0083    0.0083    0.0447    0.0447    1.0000
    0.0093    0.0093    0.0093    0.0093    0.0430    0.0430    1.0000
    0.0103    0.0103    0.0103    0.0103    0.0413    0.0413    1.0000
    0.0113    0.0113    0.0113    0.0113    0.0397    0.0397    1.0000
    0.0123    0.0123    0.0123    0.0123    0.0380    0.0380    1.0000
    0.0133    0.0133    0.0133    0.0133    0.0364    0.0364    1.0000
    0.0143    0.0143    0.0143    0.0143    0.0347    0.0347    1.0000
    0.0153    0.0153    0.0153    0.0153    0.0330    0.0330    1.0000
    0.0163    0.0163    0.0163    0.0163    0.0314    0.0314    1.0000
    0.0173    0.0173    0.0173    0.0173    0.0297    0.0297    1.0000
    0.0183    0.0183    0.0183    0.0183    0.0281    0.0281    1.0000
    0.0193    0.0193    0.0193    0.0193    0.0264    0.0264    1.0000
    0.0203    0.0203    0.0203    0.0203    0.0247    0.0247    1.0000
    0.0213    0.0213    0.0213    0.0213    0.0231    0.0231    1.0000
    0.0223    0.0223    0.0223    0.0223    0.0214    0.0214    1.0000
    0.0233    0.0233    0.0233    0.0233    0.0198    0.0198    1.0000
    0.0243    0.0243    0.0243    0.0243    0.0181    0.0181    1.0000
    0.0253    0.0253    0.0253    0.0253    0.0164    0.0164    1.0000
    0.0263    0.0263    0.0263    0.0263    0.0148    0.0148    1.0000
    0.0273    0.0273    0.0273    0.0273    0.0131    0.0131    1.0000
    0.0283    0.0283    0.0283    0.0283    0.0115    0.0115    1.0000
    0.0293    0.0293    0.0293    0.0293    0.0098    0.0098    1.0000
    0.0303    0.0303    0.0303    0.0303   -0.0569   -0.0569   -0.1122
    0.0010    0.0010    0.0010    0.0010         0         0    1.0000
    0.0020    0.0020    0.0020    0.0020         0         0    1.0000
    0.0030    0.0030    0.0030    0.0030         0         0    1.0000
    0.0040    0.0040    0.0040    0.0040         0         0    1.0000
    0.0050    0.0050    0.0050    0.0050         0         0    1.0000
    0.0060    0.0060    0.0060    0.0060         0         0    1.0000
    0.0070    0.0070    0.0070    0.0070         0         0    1.0000
    0.0080    0.0080    0.0080    0.0080         0         0    1.0000
    0.0090    0.0090    0.0090    0.0090         0         0    1.0000
    0.0100    0.0100    0.0100    0.0100         0         0    1.0000
    0.0110    0.0110    0.0110    0.0110         0         0    1.0000
    0.0120    0.0120    0.0120    0.0120         0         0    1.0000
    0.0130    0.0130    0.0130    0.0130         0         0    1.0000
    0.0140    0.0140    0.0140    0.0140         0         0    1.0000
    0.0150    0.0150    0.0150    0.0150         0         0    1.0000
    0.0160    0.0160    0.0160    0.0160         0         0    1.0000
    0.0170    0.0170    0.0170    0.0170         0         0    1.0000
    0.0180    0.0180    0.0180    0.0180         0         0    1.0000
    0.0190    0.0190    0.0190    0.0190         0         0    1.0000
    0.0200    0.0200    0.0200    0.0200         0         0    1.0000
    0.0210    0.0210    0.0210    0.0210         0         0    1.0000

| | | | | | | | | | | | |
|---|---|---|---|---|---|---|---|---|---|---|---|
| 0.0220 | 0.0220 | 0.0220 | 0.0220 | 0 | 0 | 0.0220 | 0.0220 | 0.0220 | 0.0220 | 0 | 0 |
| 1.0000 | | | | | | 1.0000 | | | | | |
| 0.0230 | 0.0230 | 0.0230 | 0.0230 | 0 | 0 | 0.0230 | 0.0230 | 0.0230 | 0.0230 | 0 | 0 |
| 1.0000 | | | | | | 1.0000 | | | | | |
| 0.0240 | 0.0240 | 0.0240 | 0.0240 | 0 | 0 | 0.0240 | 0.0240 | 0.0240 | 0.0240 | 0 | 0 |
| 1.0000 | | | | | | 1.0000 | | | | | |
| 0.0250 | 0.0250 | 0.0250 | 0.0250 | 0 | 0 | 0.0250 | 0.0250 | 0.0250 | 0.0250 | 0 | 0 |
| 1.0000 | | | | | | 1.0000 | | | | | |
| 0.0260 | 0.0260 | 0.0260 | 0.0260 | 0 | 0 | 0.0260 | 0.0260 | 0.0260 | 0.0260 | 0 | 0 |
| 1.0000 | | | | | | 1.0000 | | | | | |
| 0.0270 | 0.0270 | 0.0270 | 0.0270 | 0 | 0 | 0.0270 | 0.0270 | 0.0270 | 0.0270 | 0 | 0 |
| 1.0000 | | | | | | 1.0000 | | | | | |
| 0.0280 | 0.0280 | 0.0280 | 0.0280 | 0 | 0 | 0.0280 | 0.0280 | 0.0280 | 0.0280 | 0 | 0 |
| 1.0000 | | | | | | 1.0000 | | | | | |
| 0.0290 | 0.0290 | 0.0290 | 0.0290 | 0 | 0 | 0.0290 | 0.0290 | 0.0290 | 0.0290 | 0 | 0 |
| 1.0000 | | | | | | 1.0000 | | | | | |
| 0.0300 | 0.0300 | 0.0300 | 0.0300 | 0 | 0 | 0.0300 | 0.0300 | 0.0300 | 0.0300 | 0 | 0 |
| 1.0000 | | | | | | 1.0000 | | | | | |
| 0.0010 | 0.0010 | 0.0010 | 0.0010 | 0 | 0 | 0.0010 | 0.0010 | 0.0010 | 0.0010 | 0 | 0 |
| 1.0000 | | | | | | 1.0000 | | | | | |
| 0.0020 | 0.0020 | 0.0020 | 0.0020 | 0 | 0 | 0.0020 | 0.0020 | 0.0020 | 0.0020 | 0 | 0 |
| 1.0000 | | | | | | 1.0000 | | | | | |
| 0.0030 | 0.0030 | 0.0030 | 0.0030 | 0 | 0 | 0.0030 | 0.0030 | 0.0030 | 0.0030 | 0 | 0 |
| 1.0000 | | | | | | 1.0000 | | | | | |
| 0.0040 | 0.0040 | 0.0040 | 0.0040 | 0 | 0 | 0.0040 | 0.0040 | 0.0040 | 0.0040 | 0 | 0 |
| 1.0000 | | | | | | 1.0000 | | | | | |
| 0.0050 | 0.0050 | 0.0050 | 0.0050 | 0 | 0 | 0.0050 | 0.0050 | 0.0050 | 0.0050 | 0 | 0 |
| 1.0000 | | | | | | 1.0000 | | | | | |
| 0.0060 | 0.0060 | 0.0060 | 0.0060 | 0 | 0 | 0.0060 | 0.0060 | 0.0060 | 0.0060 | 0 | 0 |
| 1.0000 | | | | | | 1.0000 | | | | | |
| 0.0070 | 0.0070 | 0.0070 | 0.0070 | 0 | 0 | 0.0070 | 0.0070 | 0.0070 | 0.0070 | 0 | 0 |
| 1.0000 | | | | | | 1.0000 | | | | | |
| 0.0080 | 0.0080 | 0.0080 | 0.0080 | 0 | 0 | 0.0080 | 0.0080 | 0.0080 | 0.0080 | 0 | 0 |
| 1.0000 | | | | | | 1.0000 | | | | | |
| 0.0090 | 0.0090 | 0.0090 | 0.0090 | 0 | 0 | 0.0090 | 0.0090 | 0.0090 | 0.0090 | 0 | 0 |
| 1.0000 | | | | | | 1.0000 | | | | | |
| 0.0100 | 0.0100 | 0.0100 | 0.0100 | 0 | 0 | 0.0100 | 0.0100 | 0.0100 | 0.0100 | 0 | 0 |
| 1.0000 | | | | | | 1.0000 | | | | | |
| 0.0110 | 0.0110 | 0.0110 | 0.0110 | 0 | 0 | 0.0110 | 0.0110 | 0.0110 | 0.0110 | 0 | 0 |
| 1.0000 | | | | | | 1.0000 | | | | | |
| 0.0120 | 0.0120 | 0.0120 | 0.0120 | 0 | 0 | 0.0120 | 0.0120 | 0.0120 | 0.0120 | 0 | 0 |
| 1.0000 | | | | | | 1.0000 | | | | | |
| 0.0130 | 0.0130 | 0.0130 | 0.0130 | 0 | 0 | 0.0130 | 0.0130 | 0.0130 | 0.0130 | 0 | 0 |
| 1.0000 | | | | | | 1.0000 | | | | | |
| 0.0140 | 0.0140 | 0.0140 | 0.0140 | 0 | 0 | 0.0140 | 0.0140 | 0.0140 | 0.0140 | 0 | 0 |
| 1.0000 | | | | | | 1.0000 | | | | | |
| 0.0150 | 0.0150 | 0.0150 | 0.0150 | 0 | 0 | 0.0150 | 0.0150 | 0.0150 | 0.0150 | 0 | 0 |
| 1.0000 | | | | | | 1.0000 | | | | | |
| 0.0160 | 0.0160 | 0.0160 | 0.0160 | 0 | 0 | 0.0160 | 0.0160 | 0.0160 | 0.0160 | 0 | 0 |
| 1.0000 | | | | | | 1.0000 | | | | | |
| 0.0170 | 0.0170 | 0.0170 | 0.0170 | 0 | 0 | 0.0170 | 0.0170 | 0.0170 | 0.0170 | 0 | 0 |
| 1.0000 | | | | | | 1.0000 | | | | | |
| 0.0180 | 0.0180 | 0.0180 | 0.0180 | 0 | 0 | 0.0180 | 0.0180 | 0.0180 | 0.0180 | 0 | 0 |
| 1.0000 | | | | | | 1.0000 | | | | | |
| 0.0190 | 0.0190 | 0.0190 | 0.0190 | 0 | 0 | 0.0190 | 0.0190 | 0.0190 | 0.0190 | 0 | 0 |
| 1.0000 | | | | | | 1.0000 | | | | | |
| 0.0200 | 0.0200 | 0.0200 | 0.0200 | 0 | 0 | 0.0200 | 0.0200 | 0.0200 | 0.0200 | 0 | 0 |
| 1.0000 | | | | | | 1.0000 | | | | | |
| 0.0210 | 0.0210 | 0.0210 | 0.0210 | 0 | 0 | 0.0210 | 0.0210 | 0.0210 | 0.0210 | 0 | 0 |
| 1.0000 | | | | | | 1.0000 | | | | | |

| | | | | | |
|---|---|---|---|---|---|
| 0.0220 | 0.0220 | 0.0220 | 0.0220 | 0 | 0 |
| 1.0000 | | | | | |
| 0.0230 | 0.0230 | 0.0230 | 0.0230 | 0 | 0 |
| 1.0000 | | | | | |
| 0.0240 | 0.0240 | 0.0240 | 0.0240 | 0 | 0 |
| 1.0000 | | | | | |
| 0.0250 | 0.0250 | 0.0250 | 0.0250 | 0 | 0 |
| 1.0000 | | | | | |
| 0.0260 | 0.0260 | 0.0260 | 0.0260 | 0 | 0 |
| 1.0000 | | | | | |
| 0.0270 | 0.0270 | 0.0270 | 0.0270 | 0 | 0 |
| 1.0000 | | | | | |
| 0.0280 | 0.0280 | 0.0280 | 0.0280 | 0 | 0 |
| 1.0000 | | | | | |
| 0.0290 | 0.0290 | 0.0290 | 0.0290 | 0 | 0 |
| 1.0000 | | | | | |
| 0.0300 | 0.0300 | 0.0300 | 0.0300 | 0 | 0 |
| 1.0000 | | | | | |

Where $1^{st}$ and $2^{nd}$ coloumn indicates the initial swarm position and velocity, $3^{rd}$ and $4^{th}$ column represents the linear and angular position and velocity and $5^{th}$ and $6^{th}$ column represents updated and the $7^{th}$ column represents the objective function values. The following Figure 2 represents the plot which is drawn between the Swarm position and its objective function values.

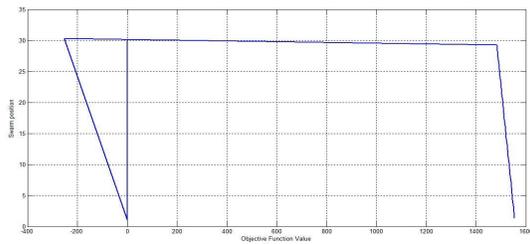

**Fig. 2** ES with PSO swarm position vs objective function values

The following Figure 3 represents the swarm motion in 3D-plot with ESPSO optimization methods for performance evolution in BLDC motor.

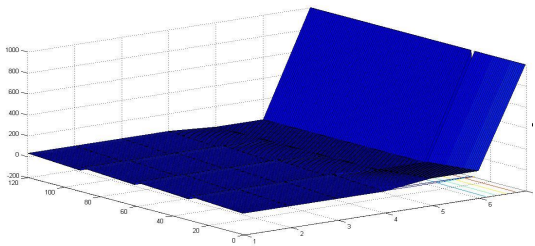

**Fig. 3** Swarm movement in mesh plot with ES-PSO

The final iterative simulation results with ES with Firefly algorithm is as listed below.

```
   firefly =
     1.0e+003 *
       0.0010    0.0010   -0.0000    0.0010
       0.0020    0.0020   -0.0000    0.0020
       0.0030    0.0030   -0.0000    0.0030
       0.0040    0.0040   -0.0000    0.0040
       0.0050    0.0050   -0.0000    0.0050
       0.0060    0.0060   -0.0000    0.0060
       0.0070    0.0070   -0.0000    0.0070
       0.0080    0.0080   -0.0000    0.0080
       0.0090    0.0090   -0.0000    0.0090
       0.0100    0.0100   -0.0000    0.0100
       0.0110    0.0110   -0.0000    0.0110
       0.0120    0.0120   -0.0000    0.0120
       0.0130    0.0130   -0.0000    0.0130
       0.0140    0.0140   -0.0000    0.0140
       0.0150    0.0150   -0.0000    0.0150
       0.0160    0.0160   -0.0000    0.0160
       0.0170    0.0170   -0.0000    0.0170
       0.0180    0.0180   -0.0000    0.0180
       0.0190    0.0190   -0.0000    0.0190
       0.0200    0.0200   -0.0000    0.0200
       0.0210    0.0210    0.0210    0.0210
       0.0220    0.0220    0.0220    0.0220
       0.0230    0.0230    0.0230    0.0230
       0.0240    0.0240    0.0240    0.0240
       0.0250    0.0250    0.0250    0.0250
       0.0001    0.0007    0.0006    1.0000
       0.0002    0.0006    0.0006    1.0000
       0.0002    0.0005    0.0006    1.0000
       0.0004    0.0005    0.0002    1.0000
       0.0000    0.0003    0.0003    1.0000

   firefly =

     1.0e+003 *

       0.0010    0.0010   -0.0000    0.0010
       0.0020    0.0020   -0.0000    0.0020
       0.0030    0.0030   -0.0000    0.0030
       0.0040    0.0040   -0.0000    0.0040
       0.0050    0.0050   -0.0000    0.0050
       0.0060    0.0060   -0.0000    0.0060
       0.0070    0.0070   -0.0000    0.0070
       0.0080    0.0080   -0.0000    0.0080
       0.0090    0.0090   -0.0000    0.0090
       0.0100    0.0100   -0.0000    0.0100
       0.0110    0.0110   -0.0000    0.0110
       0.0120    0.0120   -0.0000    0.0120
       0.0130    0.0130   -0.0000    0.0130
       0.0140    0.0140   -0.0000    0.0140
       0.0150    0.0150   -0.0000    0.0150
       0.0160    0.0160   -0.0000    0.0160
       0.0170    0.0170   -0.0000    0.0170
       0.0180    0.0180   -0.0000    0.0180
       0.0190    0.0190   -0.0000    0.0190
       0.0200    0.0200    0.0201    0.0200
       0.0210    0.0210    0.0210    0.0210
       0.0220    0.0220    0.0220    0.0220
       0.0230    0.0230    0.0230    0.0230
       0.0240    0.0240    0.0240    0.0240
       0.0250    0.0250    0.0250    0.0250
       0.0001    0.0007    0.0006    1.0000
       0.0002    0.0006    0.0006    1.0000
       0.0002    0.0005    0.0006    1.0000
```

```
0.0004   0.0005   0.0002   1.0000
0.0000   0.0003   0.0003   1.0000
```

Here 1st and 2nd column consists of positions of Firefly ith and jth positions and 3rd coumn consists of update position of ith firefly and 4th column consists of objective function values.

The following Figure 4 represents the plot which is drawn between the Firefly ith and jth positions. The following Figure 5 represents the Firefly motion in 3D-plot with ESFFA optimization methods for performance evolution in BLDC motor.

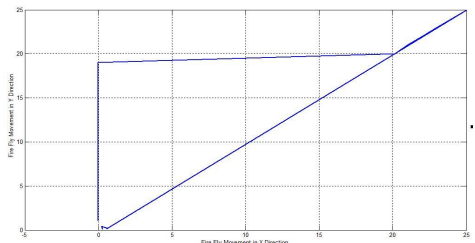

**Fig. 4** ES - FFA Firefly Trajectories of ith position vs jth position

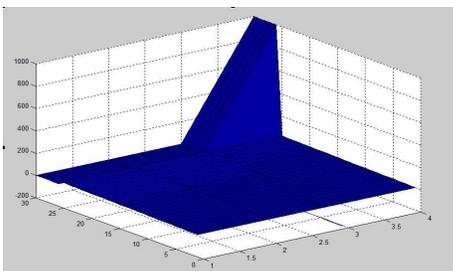

**Fig. 5** Fireflies movement in mesh plot with ES-FFA

**Simulation Validation With Assumed Parameters**
Parameters in ES
Levys walking step length( $s$ ) = 5
$\lambda = 1.5$
Parameters in PSO
Max. No. of Iterations=20;
No. of Swarms=30;
Correction factors $c_1, c_2 = 2$
Weight of the ith swarm( $w_i$ ) = 1

Parameters in FFA
Max. No. of Iterations=20;
No. of Fireflies=30;
Absorption coefficient ( $\gamma$ )=1
Initial attractiveness constant( $\beta_0$ )=0.2

Correction factor which accelerates the optimization speed by controlling the Firefly motion( $\alpha$ ) =0.3
Initial assigned global objective function value=1000

## CONCLUSION

In this paper, comparison among Brushless motor with other conventional motors were discussed. For the performance evolution of brushless motor an adaptive system is developed. But the noise signals and the parametric variations mainly effects the stability of the brushless motors under practical approach. The stability of the system under closed loop operation is tested with Lyapunov stability theorems. But the stability is guaranteed only if the powerful and efficient optimization techniques are adapted to the closed loop operation of brushless motors. Hence, ES with Levy's walk distribution exhibits optimum diversified global search when compared to other search algorithms. Then for the intensified local search operation PSO and FFA are implemented. The simulation results are justified that the combination of ES with PSO or FFA shows effectiveness in optimization.

## REFERENCES

[1] Subhash Challa, Mark R. Morelande, Darko Muscki, Robin J. Evans. "*Fundamentals Of Object Tracking,*" 1st ed., Cambridge University Press, New York: 2011.
[2] H. J. Jin, J. M. Hwang, and J. M. Lee, "*A balancing control strategy for a one wheel pendulum robot based on dynamic model decomposition: simulation and experiments*," IEEE/ASME Trans. On Mechatronics, vol. 16, no. 4, pp. 763-768, 2011.
[3] C. C. Tsai, H. C. Huang, and S. C. Lin, "*Adaptive neural network control of self-balancing two-wheeled scooter*," IEEE Trans. on Industrial Electronics, vol. 57, no. 4, pp. 1420-1428, 2010.
[4] TJE Miller, "*Brushless Permanent-Magnet and Reluctance Motor Drives*", Oxford Science Publications:1989.
[5] C. H. Huang, W. J. Wang, and C. H. Chiu, "*Design and implementation of fuzzy control on a two-wheel inverted pendulum system*," IEEE Trans. on Industrial Electronics, vol. 58, no. 7, pp. 2988-3001, 2011.
[6] V. Amerongen and J. Intelligent, "*Control (Part 1)-MRAS, lecture notes,*" University of Twente, The Netherlands, March 2004.
[7] Appalabathula Venkatesh, Dr. G. Raja Rao, " *An Improved Adaptive Control System for a Two-Wheel Inverted Pendulum-Mobile Robot using Eagle Strategy with a Particle Swarm Optimization*," Journal of Emerging Technologies and Innovative Research, vol. 5,issue 7, July.2018.
[8] Hamza Yapici, Nurettin Cetinkaya, "*An Improved Particle Swarm Optimization Algorithm Using Eagle Strategy for Power Loss Minimization*" Mathematical Problems in Engineering, vol. 2017.
[9] N. D. Cuong, N. Van Lanh, and D. Van Huyen, "*Design of MRAS-based adaptive control systems,*" in Proc. IEEE 2013 International Conference on Control, Automation and Information Sciences, pp. 79-84, 2013.
[10] Derek Atherton, "*An Introduction to Non-Linearity in Control System*": 2011.
[11] Nguyen Duy Cuong and Gia Thi Dinh and Tran Xuan Minh, "*Direct MRAS Based an Adaptive Control System for a Two-Wheel Mobile Robot*" Journal of Automation and Control Engineering ,Vol.3,No.3, June 2015.
[12] Xin-She Yang, Suash Deb, "*Eagle Strategy Using Levy walk and Firefly Algorithms for Stochastic Optimization*" Natures Inspired Coorperative Strategies for Optimization (NICSO 2010), pp. 101-111: 2010.
[13] Appalabathula Venkatesh, Shankar Nalinakshan, S S Kiran, Pradeepa H, "*Energy transmission control for a Grid connected modern power system Non-Linear loads with a Series Multi-Stage Transformer Voltage Reinjection with controlled converters*" International Journal of Engineering Trends and Technology, Vol. 68,Issue 8, August 2020.